\documentclass[preprint,preprintnumbers]{revtex4-1}
\usepackage{amssymb, amsthm}
\usepackage{amsmath}    
\usepackage{mathrsfs}   
\usepackage{graphicx}   
\usepackage{color}      
\usepackage{footnote}   

\usepackage{slashed}    
\usepackage{subfigure}  
\usepackage{hyperref}   

\usepackage{rotating}   
\usepackage{multirow}   

\begin{document}

\preprint{NT@UW-17-05}
\title{Validity of the Weizs\"{a}cker-Williams Approximation and the Analysis of Beam Dump Experiments: Production of an axion, a dark photon, or a new axial-vector boson}
\author{Yu-Sheng Liu}\email{mestelqure@gmail.com}
\author{Gerald A. Miller}\email{miller@phys.washington.edu}
\affiliation{Department of Physics,	University of Washington, Seattle, Washington 98195-1560, U.S.A.}
\date{\today}

\begin{abstract}
Beam dump experiments have been used to search for new particles, $\phi$, with null results interpreted in terms of limits on masses $m_\phi$ and coupling constants $\epsilon$. However these limits have been obtained by using approximations [including the Weizs\"{a}cker-Williams (WW) approximation] or Monte-Carlo simulations. We display methods to obtain the cross section and the resulting particle production rates without using approximations on the phase space integral or Monte-Carlo simulations. In our previous work we examined the case of the new scalar boson production; in this paper we explore all possible new spin-0 and spin-1 particles. We show that the approximations cannot be used to obtain accurate values of cross sections. The corresponding exclusion plots differ by substantial amounts when seen on a linear scale. Furthermore, a new region ($m_\phi<2m_e$) of parameter space can be explored without using one of the common approximations, $m_\phi\gg m_e$. We derive new expressions for the three photon decays of dark photon and four photon decays of new axial-vector bosons. As a result, the production cross section and exclusion region of different low mass ($m_\phi<2m_e$) bosons are very different. Moreover, our method can be used as a consistency check for Monte-Carlo simulations.
\end{abstract}
\maketitle


\section{introduction}
In our previous work \cite{Liu:2016mqv}, as an example, we used the beam dump experiment E137 \cite{Bjorken:1988as} and the production of a new scalar boson to demonstrate our technique for analyzing of beam dump experiments. In this paper, we further include all possible new spin-0 and spin-1 particles, which we denote $\phi$; they are pseudoscalar, vector, and axial-vector bosons.

Beam dump experiments have been aimed at searching for new particles, such as dark photons and axions (see, e.g. \cite{Essig:2013lka} and references therein) that decay to lepton pairs and/or photons. Electron beam dumps in particular have received a large amount of theoretical attention in recent years~\cite{Bjorken:2009mm,Andreas:2012mt}. The typical setup of an electron beam dump experiment is to dump an electron beam into a target, in which the electrons are stopped. The new particles produced by the bremsstrahlung-like process pass through a shield region and decay. These new particles can be detected by their decay products, electron and/or photon pairs, measured by the detector downstream of the decay region. Previous earlier work simplified the necessary phase space integral by using the Weizs\"{a}cker-Williams (WW) approximation \cite{vonWeizsacker:1934nji,Williams:1935dka} which, also known as method of virtual quanta, is a semiclassical approximation. The idea is that the electromagnetic field generated by a fast moving charged particle is nearly transverse which is like a plane wave and can be approximated by real photon. The use of the  WW approximation in bremsstrahlung processes was developed in Refs.~\cite{Kim:1973he,Tsai:1973py} and applied to beam dump experiments in Refs.~\cite{Bjorken:2009mm,Tsai:1986tx}. The WW approximation simplifies evaluation of the integral over phase space and approximates the 2 particle to 3 particle (2 to 3) cross section in terms of a 2 particle to 2 particle (2 to 2) cross section. For the WW approximation to work in a beam dump experiment, it needs the incoming beam energy to be much greater than the mass of the new particle, $m_\phi$, and electron mass $m_e$. 

The previous work \cite{Bjorken:2009mm} used the following three approximations:
\begin{enumerate}
\item WW approximation;
\item a further simplification of the phase space integral, see Eq. (\ref{eq:tmin tmax});
\item $m_\phi\gg m_e$.
\end{enumerate}
The combination of the first two approximations has been denoted \cite{Kim:1973he} the improved WW (IWW) approximation. The name ``improved WW" might be somewhat misleading since the procedure reduces the computational time but does not improve accuracy). In this paper, we will focus on examining the validity of WW and IWW approximations for the production of axions, dark photons and new axial-vector bosons. The third approximation used to simplify the calculation of amplitude, however, is not in our scope because it is merely a special case by cutting off our results when $m_\phi\lesssim 2m_e$. Nevertheless, we should point out that without using the third approximation we can use beam dump experiments to explore a larger parameter space.

The outline of this paper is as follows. In Sec.~\ref{sec:dynamics}, we setup the dynamics, and then calculate the decay width of new particles and the squared amplitude for 2 to 3 and 2 to 2 processes. In Sec. \ref{sec:cross section}, we show the 2 to 3 cross sections in the lab frame without any approximation on phase space and WW approximation is discussed. In Sec.~\ref{sec:cross section comparison}, we derive and compare the cross sections with and without approximations. In Sec.~\ref{sec:particle production}, we discuss the number of new particles produced in beam dump experiments. In Sec.~\ref{sec:exclusion plots}, we compare the exclusion plots for different bosons with and without approximations. A discussion is presented in Sec.~\ref{sec:discussion}.

\section{dynamics}\label{sec:dynamics}
For simplicity, we assume that there is only one new boson $\phi$, which only couples to electron by a Yukawa interaction, i.e. the boson does not couple to other standard model fermions other than electron. The Lagrangian contains either one of the following interactions
\begin{align}
\mathcal{L}_P&=i e\epsilon_P\phi\bar\psi\gamma_5\psi\nonumber\\
\mathcal{L}_V&=e\epsilon_V\phi_\mu\bar\psi\gamma^\mu\psi\\
\mathcal{L}_A&=e\epsilon_A\phi_\mu\bar\psi\gamma_5\gamma^\mu\psi\nonumber
\end{align}
where the subscripts $P$, $V$, and $A$ correspond to pseudoscalar, vector, and axial-vector, respectively; $\epsilon=g/e$, $g$ is the coupling of the new boson, and $e$ is the electric charge; $\psi$ is the electron field; $\gamma_5=-\frac{i}{4!}\epsilon_{\mu\nu\rho\sigma}\gamma^\mu\gamma^\nu\gamma^\rho\gamma^\sigma$; we choose the convention that there is an extra $i$ in $\mathcal{L}_P$, such that $\epsilon_P$ can be a non-negative number.

If $m_\phi>2m_e$, the dominant new boson decay is to electron pairs
\begin{align}\label{eq:decay to electrons}
\Gamma_{P}(\phi\to e^+e^-)&=\epsilon_P^2\frac{\alpha}{2}m_\phi\left(1-\frac{4m_e^2}{m_\phi^2}\right)^{1/2}\nonumber\\
\Gamma_{V}(\phi\to e^+e^-)&=\epsilon_V^2\frac{\alpha}{3}m_\phi\left(1+\frac{2m_e^2}{m_\phi^2}\right)\left(1-\frac{4m_e^2}{m_\phi^2}\right)^{1/2}\\
\Gamma_{A}(\phi\to e^+e^-)&=\epsilon_A^2\frac{\alpha}{3}m_\phi\left(1-\frac{4m_e^2}{m_\phi^2}\right)^{3/2}\nonumber,
\end{align}
where $\alpha$ is the fine structure constant.

If $m_\phi<2m_e$, the dominant decay channel involves photons produced through the electron loop. For pseudoscalar, it decays to two photons 
\begin{align}\label{eq:P decay to photons}
\Gamma_{P}(\phi\to\gamma\gamma)=\epsilon_P^2\frac{\alpha^3}{4\pi^2}\frac{m_\phi^3}{m_e^2}f_P\left(\frac{m_\phi^2}{4m_e^2}\right)
\end{align}
where $f_P(\tau)=\frac{1}{64\tau^2}\left|\ln\left[1-2\left(\tau+\sqrt{\tau^2-\tau}\right)\right]^2 \right|^2$. For spin-1 particles, however, the two photon decay channel is forbidden by Landau--Yang theorem \cite{Landau:1948kw,Yang:1950rg,Zhemchugov:2014dza}. Therefore, the dominant decay channel of the vector boson is 3 photon decay 
\begin{align}
\Gamma(\phi\to\gamma_1+\gamma_1+\gamma_3)=\frac{1}{64S\pi^3 m_\phi}\int^\frac{m_\phi}{2}_0dE_1\int^\frac{m_\phi}{2}_{\frac{m_\phi}{2}-E1}dE_2|\mathcal{M}|^2
\end{align}
where $S$ is the symmetry factor accounting for identical particles in the final state and in this case $S=3!$; $E_1$ and $E_2$ are energy of $\gamma_1$ and $\gamma_2$, respectively; $\mathcal{M}$ is the amplitude containing 6 diagrams. We express the decay rate in term of $\frac{m_\phi}{m_e}$ expansion
\begin{align}\label{eq:V decay to photons}
\Gamma_{V}(\phi\to3\gamma)=\epsilon_V^2\frac{\alpha^4}{2^7 3^6 5^2 \pi^3}\frac{m_\phi^9}{m_e^8}\left[\frac{17}{5}+\frac{67}{42}\frac{m_\phi^2}{m_e^2}+\frac{128941}{246960}\frac{m_\phi^4}{m_e^4}+\mathcal{O}\left(\frac{m_\phi^6}{m_e^6}\right)\right].
\end{align}
The leading term of this result agrees with \cite{Pospelov:2008jk}, which used effective field theory.

For axial-vector, the 3 photon decay channel is further forbidden by the charge conjugation symmetry (similar with the argument of Furry's theorem). Thus the dominant decay channel of the axial-vector boson is 4 photon decay. There are 24 diagrams and the 4 body phase space integral of the decay rate is done in Refs. \cite{Anastasiou:2003gr,Asatrian:2012tp}. We express the result in term of $\frac{m_\phi}{m_e}$ expansion
\begin{align}\label{eq:A decay to photons}
\Gamma_{A}(\phi\to 4\gamma)=\epsilon_A^2\frac{127\alpha^5}{2^{11} 3^8 5^4 7^2 \pi^4}\frac{m_\phi^{13}}{m_e^{12}}+\mathcal{O}\left(\frac{m_\phi^{15}}{m_e^{14}}\right).
\end{align}

\subsection{2 to 3 production}
\begin{figure}
\centering
\includegraphics[scale=0.8]{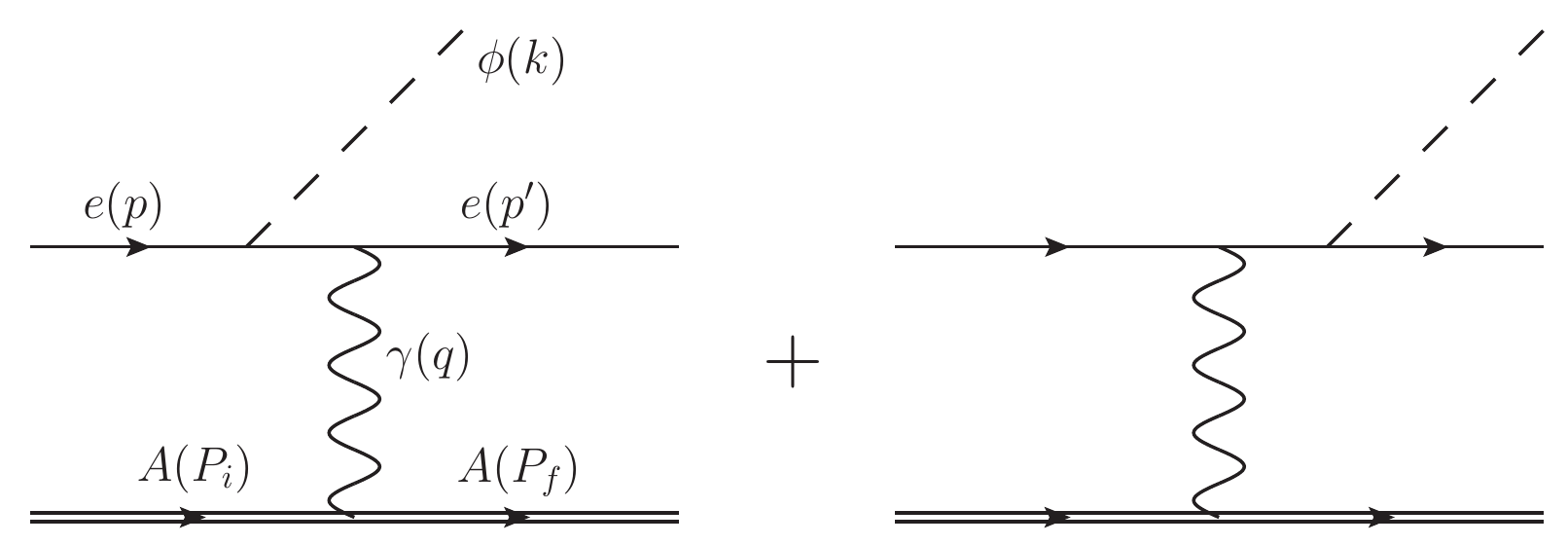}
\caption{\label{fig:2 to 3} Lowest order 2 to 3 production process: $e(p)+A(P_i)\rightarrow e(p')+A(P_f)+\phi(k)$. $A$, $\gamma$, $e$, and $\phi$ stand for the target atom, photon, electron, and the new boson.}
\end{figure}

The leading production process is the bremsstrahlung-like radiation of the new particle from the electron, shown in Fig.~\ref{fig:2 to 3},
\begin{align}\label{eq:2 to 3 production process}
e(p)+A(P_i)\rightarrow e(p')+A(P_f)+\phi(k)
\end{align}
where $e$, $A$, and $\phi$ stand for electron, target atom, and the new particle, respectively. We define the following quantities using the mostly-plus metric
\begin{align}\label{eq:2 to 3 variables}
\tilde{s}&=-(p'+k)^2-m_e^2=-2p'\cdotp k+m_\phi^2\nonumber\\
\tilde{u}&=-(p-k)^2-m_e^2=2p\cdotp k+m_\phi^2\nonumber\\
t_2&=-(p'-p)^2=2p'\cdotp p+2m_e^2\\
q&=P_i-P_f\nonumber\\
t&=q^2\nonumber
\end{align}
which satisfy
\begin{align}
\tilde{s}+t_2+\tilde{u}+t=m_\phi^2.
\end{align}

For definiteness, we assume the atom is a scalar boson (its spin is not consequential here) so that the Feynman rule for the photon-atom vertex is 
\begin{align}
ieF(q^2)(P_i+P_f)_\mu\equiv ieF(q^2)P_\mu
\end{align}
where $F(q^2)$ is the form factor which accounts for the nuclear form factor \cite{DeJager:1987qc} and the atomic form factor \cite{atomic form factor}. Here, we only include the elastic form factor since the contribution of the inelastic one is much smaller and can be neglected in computing the cross section. The amplitude of the process in Fig.~\ref{fig:2 to 3} using the mostly-plus metric is
\begin{align}
\mathcal{M}^{23}_P&=ie^2g_P\frac{F(q^2)}{q^2}\bar{u}_{p',s'}\left[\slashed{P}\frac{(\slashed{p}-\slashed{k})-m_e}{\tilde{u}}\gamma_5+\gamma_5\frac{(\slashed{p'}+\slashed{k})-m_e}{\tilde{s}}\slashed{P}\right]u_{p,s}\nonumber\\
\mathcal{M}^{23}_V&=e^2g_V\frac{F(q^2)}{q^2}\tilde{\epsilon}^\mu_{k,\lambda}\bar{u}_{p',s'}\left[\slashed{P}\frac{(\slashed{p}-\slashed{k})-m_e}{\tilde{u}}\gamma_\mu+\gamma_\mu\frac{(\slashed{p'}+\slashed{k})-m_e}{\tilde{s}}\slashed{P}\right]u_{p,s}\\
\mathcal{M}^{23}_A&=e^2g_A\frac{F(q^2)}{q^2}\tilde{\epsilon}^\mu_{k,\lambda}\bar{u}_{p',s'}\left[\slashed{P}\frac{(\slashed{p}-\slashed{k})-m_e}{\tilde{u}}\gamma_5\gamma_\mu+\gamma_5\gamma_\mu\frac{(\slashed{p'}+\slashed{k})-m_e}{\tilde{s}}\slashed{P}\right]u_{p,s}\nonumber
\end{align}
where $P$, $V$, and $A$ stand for pseudo-scalar, vector, and axial-vector, respectively; $u_{p,s}$ is the electron spinor and $s=\pm 1$; $\tilde{\epsilon}$ is the polarization of the new spin-1 particle and $\lambda=0,\,\pm 1$. The polarization sum for the new massive spin-1 particle is
\begin{align}
\sum_\lambda\tilde{\epsilon}^\mu_{k,\lambda}\tilde{\epsilon}^{\nu*}_{k,\lambda}=g^{\mu\nu}+\frac{k^\mu k^\nu}{m_\phi^2}.
\end{align}
After averaging and summing over initial and final spins, we have
\begin{align}
\overline{|\mathcal{M}^{23}_P|^2}&=\left(\frac{1}{2}\sum_s\right)\sum_{s'}|\mathcal{M}_P^{23}|^2=e^4g_P^2\frac{F(q^2)^2}{q^4}\mathcal{A}^{23}_P\nonumber\\
\overline{|\mathcal{M}^{23}_{V,A}|^2}&=\left(\frac{1}{2}\sum_s\right)\sum_{s'}\sum_{\lambda}|\mathcal{M}_{V,A}^{23}|^2=e^4g_{V,A}^2\frac{F(q^2)^2}{q^4}\mathcal{A}^{23}_{V,A}
\end{align}
where
\begin{align}
\mathcal{A}^{23}_P=&-\frac{(\tilde{s}+\tilde{u})^2}{\tilde{s}\tilde{u}}P^2-\frac{4t}{\tilde{s}\tilde{u}}(P\cdotp k)^2-\frac{(\tilde{s}+\tilde{u})^2}{\tilde{s}^2\tilde{u}^2}m_\phi^2\left[P^2 t+4\left(\frac{\tilde{u}P\cdotp p+\tilde{s}P\cdotp p'}{\tilde{s}+\tilde{u}}\right)^2\right]\nonumber\\
\mathcal{A}^{23}_V=&-2\frac{\tilde{s}^2+\tilde{u}^2}{\tilde{s}\tilde{u}}P^2-\frac{8t}{\tilde{s}\tilde{u}}\left[(P\cdotp p)^2+(P\cdotp p')^2-\frac{t_2+m_\phi^2}{2}P^2\right]\nonumber\\
&-2\frac{(\tilde{s}+\tilde{u})^2}{\tilde{s}^2\tilde{u}^2}(m_\phi^2+2m_e^2)\left[P^2 t+4\left(\frac{\tilde{u}P\cdotp p+\tilde{s}P\cdotp p'}{\tilde{s}+\tilde{u}}\right)^2\right]\\
\mathcal{A}^{23}_A=&-2\frac{\tilde{s}^2+\tilde{u}^2}{\tilde{s}\tilde{u}}P^2-\frac{8t}{\tilde{s}\tilde{u}}\left[(P\cdotp p)^2+(P\cdotp p')^2-\frac{t_2-m_\phi^2}{2}P^2\right]-4m_e^2\frac{(\tilde{s}+\tilde{u})^2 P^2+4t(P\cdotp k)^2}{m_\phi^2\tilde{s}\tilde{u}}\nonumber\\
&-2\frac{(\tilde{s}-\tilde{u})^2}{\tilde{s}^2\tilde{u}^2}(m_\phi^2-4m_e^2)\left[P^2 t+4\left(\frac{\tilde{u}P\cdotp p+\tilde{s}P\cdotp p'}{\tilde{s}-\tilde{u}}\right)^2\right]\nonumber.
\end{align}

\subsection{2 to 2 production}
\begin{figure}
\centering
\includegraphics[scale=0.8]{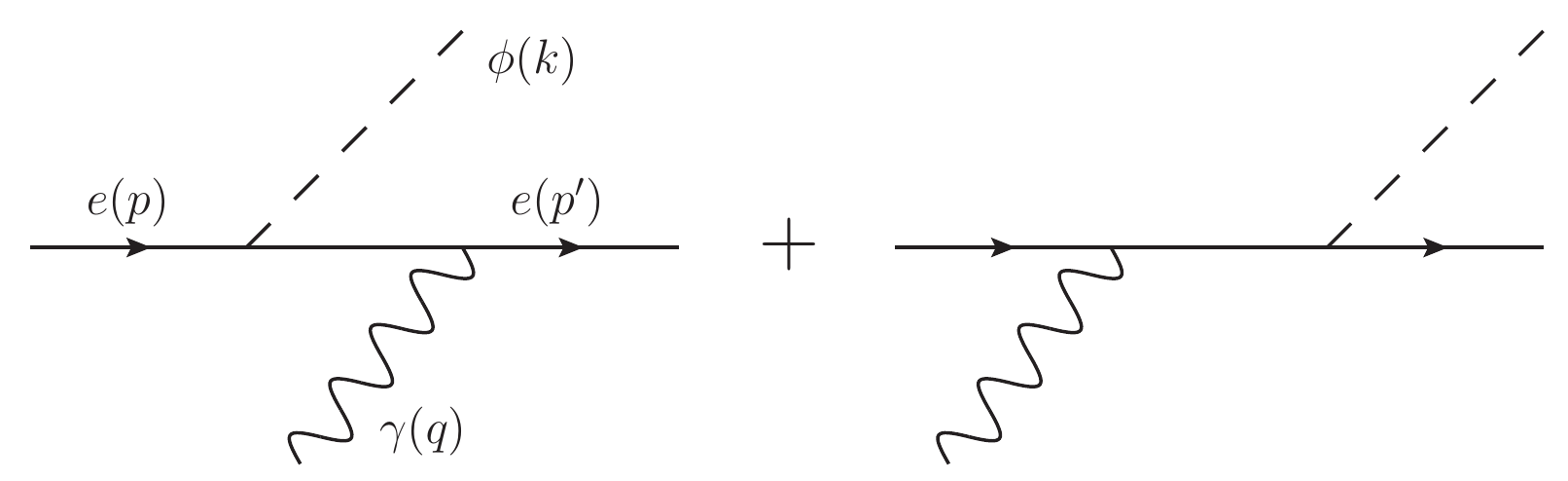}
\caption{\label{fig:2 to 2}  Lowest order 2 to 2 production process: $e(p)+\gamma(q)\rightarrow e(p')+\phi(k)$. $\gamma$, $e$, and $\phi$ stand for photon, electron, and the new boson.}
\end{figure}

For the 2 to 2 process in Fig. \ref{fig:2 to 2}, a ``subprocess'' of the full 2 to 3 interaction,
\begin{align}\label{eq:2 to 2 production process}
e(p)+\gamma(q)\rightarrow e(p')+\phi(k).
\end{align} 
With the same definition in Eq. (\ref{eq:2 to 3 variables}), $\tilde{s}$, $\tilde{u}$, and $t_2$ satisfy
\begin{align}
\tilde{s}+t_2+\tilde{u}&=m_\phi^2
\end{align}
and the amplitude in Fig. \ref{fig:2 to 2} is
\begin{align}
\mathcal{M}^{22}_P=&ieg_P\epsilon^\mu_{q,\lambda}\bar{u}_{p',s'}\left[\gamma_\mu\frac{(\slashed{p}-\slashed{k})-m_e}{\tilde{u}}\gamma_5+\gamma_5\frac{(\slashed{p'}+\slashed{k})-m_e}{\tilde{s}}\gamma_\mu\right]u_{p,s}\nonumber\\
\mathcal{M}^{22}_V=&eg_V\epsilon^\mu_{q,\lambda}\tilde{\epsilon}^\nu_{k,\lambda'}\bar{u}_{p',s'}\left[\gamma_\mu\frac{(\slashed{p}-\slashed{k})-m_e}{\tilde{u}}\gamma_\nu+\gamma_\nu\frac{(\slashed{p'}+\slashed{k})-m_e}{\tilde{s}}\gamma_\mu\right]u_{p,s}\\
\mathcal{M}^{22}_A=&eg_A\epsilon^\mu_{q,\lambda}\tilde{\epsilon}^\nu_{k,\lambda'}\bar{u}_{p',s'}\left[\gamma_\mu\frac{(\slashed{p}-\slashed{k})-m_e}{\tilde{u}}\gamma_5\gamma_\nu+\gamma_5\gamma_\nu\frac{(\slashed{p'}+\slashed{k})-m_e}{\tilde{s}}\gamma_\mu\right]u_{p,s}\nonumber
\end{align}
where $\epsilon$ is the photon polarization vector and $\lambda=\pm 1$. The polarization sum for photon is
\begin{align}
\sum_\lambda \epsilon^\mu_{q,\lambda}\epsilon^{\nu*}_{q,\lambda}=g^{\mu\nu}.
\end{align}

After averaging and summing over the initial and final spins and polarization,
\begin{align}\label{eq:2 to 2 M}
\overline{|\mathcal{M}^{22}|^2}_{P}&=\left(\frac{1}{2}\sum_s\right)\sum_{s'}\left(\frac{1}{2}\sum_\lambda\right)|\mathcal{M}^{22}_P|^2=e^2g_P^2\mathcal{A}^{22}_{P}\nonumber\\
\overline{|\mathcal{M}^{22}|^2}_{V,A}&=\left(\frac{1}{2}\sum_s\right)\sum_{s'}\left(\frac{1}{2}\sum_\lambda\right)\sum_{\lambda'}|\mathcal{M}^{22}_{V,A}|^2=e^2g_{V,A}^2\mathcal{A}^{22}_{V,A}
\end{align}
where
\begin{align}\label{eq:2 to 2 A}
\mathcal{A}^{22}_P=&-\frac{(\tilde{s}+\tilde{u})^2}{\tilde{s}\tilde{u}}+2m_\phi^2\left[\left(\frac{\tilde{s}+\tilde{u}}{\tilde{s}\tilde{u}}\right)^2m_e^2-\frac{t_2}{\tilde{s}\tilde{u}}\right]\nonumber\\
\mathcal{A}^{22}_V=&4-2\frac{(\tilde{s}+\tilde{u})^2}{\tilde{s}\tilde{u}}+4(m_\phi^2+2m_e^2)\left[\left(\frac{\tilde{s}+\tilde{u}}{\tilde{s}\tilde{u}}\right)^2m_e^2-\frac{t_2}{\tilde{s}\tilde{u}}\right]\\
\mathcal{A}^{22}_A=&4-\left(2+\frac{4m_e^2}{m_\phi^2}\right)\frac{(\tilde{s}+\tilde{u})^2}{\tilde{s}\tilde{u}}+4(m_\phi^2-4m_e^2)\left[\left(\frac{\tilde{s}+\tilde{u}}{\tilde{s}\tilde{u}}\right)^2m_e^2-\frac{t_2}{\tilde{s}\tilde{u}}\right]\nonumber.
\end{align}

\section{cross section and Weizs\"{a}cker-Williams approximation}\label{sec:cross section}
\subsection{2 to 3 cross section}
The cross section for the 2 to 3 process in the lab frame, see Fig. \ref{fig:2 to 3} and Ref. \cite{Liu:2016mqv} for more detail, is given by
\begin{align}\label{eq:2 to 3}
\frac{d\sigma}{dx d\cos\theta}=\epsilon^2\frac{\alpha^3}{8M^2}\frac{|\textbf{k}|E}{|\textbf{p}|V}\int^{t_{max}}_{t_{min}}dt\frac{F(t)^2}{t^2}\int_0^{2\pi}\frac{d\phi_q}{2\pi}\mathcal{A}^{23}
\end{align}
where $x\equiv E_k/E$; $M$ is the mass of the target atom; $\phi_q$ is azimuthal angles of \textbf{q} in the direction of $\mathbf{V}=\mathbf{k}-\mathbf{p}$; $V=|\textbf{V}|$ and $Q=|\mathbf{q}|$; $t(Q)=q^2=2M(\sqrt{M^2+Q^2}-M)$, $t_{max}=t(Q_+)$, $t_{min}=t(Q_-)$, and
\begin{align}
Q_\pm=\frac{V[\tilde{u}+2M(E'+E_f)]\pm(E'+E_f)\sqrt{\tilde{u}^2+4M\tilde{u}(E'+E_f)+4M^2V^2}}{2(E'+E_f)^2-2V^2}.
\end{align}

\subsection{Weizs\"{a}cker-Williams approximation}
It is explained in Ref.~\cite{Kim:1973he} that the WW approximation relies on the incoming electron energy being much greater than $m_\phi$ and $m_e$, such that the final state electron and the new boson are highly collinear. using WW approximation, the phase space integral can be approximated by
\begin{align}\label{eq:WW}
\frac{1}{8M^2}\int\frac{d\phi_q}{2\pi}\mathcal{A}^{23}\approx\frac{t-t_{min}}{2t_{min}}\mathcal{A}^{22}_{t=t_{min}}.
\end{align}
Following the discussion in Refs.~\cite{Liu:2016mqv,Bjorken:2009mm,Tsai:1986tx}, near $t=t_{min}$ (when $\mathbf{q}$ and $\mathbf{V}=\mathbf{k}-\mathbf{p}$ are collinear), we can approximate the following quantities
\begin{align}\label{eq:WW variables}
\tilde{s}&\approx-\frac{\tilde{u}}{1-x}\nonumber\\
\tilde{u}&\approx-xE^2\theta_\phi^2-m_\phi^2\frac{1-x}{x}-m_e^2 x\nonumber\\
t_2&\approx\frac{\tilde{u}x}{1-x}+m_\phi^2\\
V&\approx E(1-x)\nonumber\\
t_{min}&\approx\frac{\tilde{s}^2}{4E^2}\nonumber.
\end{align}
Using the above approximations to evaluated $\mathcal{A}^{22}$ at $t=t_{min}$, we have
\begin{align}\label{eq:2 to 2 A tmin}
\mathcal{A}^{22}_{P,t=t_{min}}\approx&\frac{x^2}{1-x}+2m_\phi^2\frac{\tilde{u}x+m_\phi^2(1-x)+m_e^2x^2}{\tilde{u}^2}\nonumber\\
\mathcal{A}^{22}_{V,t=t_{min}}\approx&2\frac{2-2x+x^2}{1-x}+4(m_\phi^2+2m_e^2)\frac{\tilde{u}x+m_\phi^2(1-x)+m_e^2x^2}{\tilde{u}^2}\\
\mathcal{A}^{22}_{A,t=t_{min}}\approx&\frac{4m^2x^2}{m_\phi^2(1-x)}+2\frac{2-2x+x^2}{1-x}+4(m_\phi^2-4m_e^2)\frac{\tilde{u}x+m_\phi^2(1-x)+m_e^2x^2}{\tilde{u}^2}\nonumber.
\end{align}

\section{cross section comparison}\label{sec:cross section comparison}

\begin{figure}
\centering
\subfigure[\;$d\sigma_P/dx$]{\includegraphics[scale=0.95]{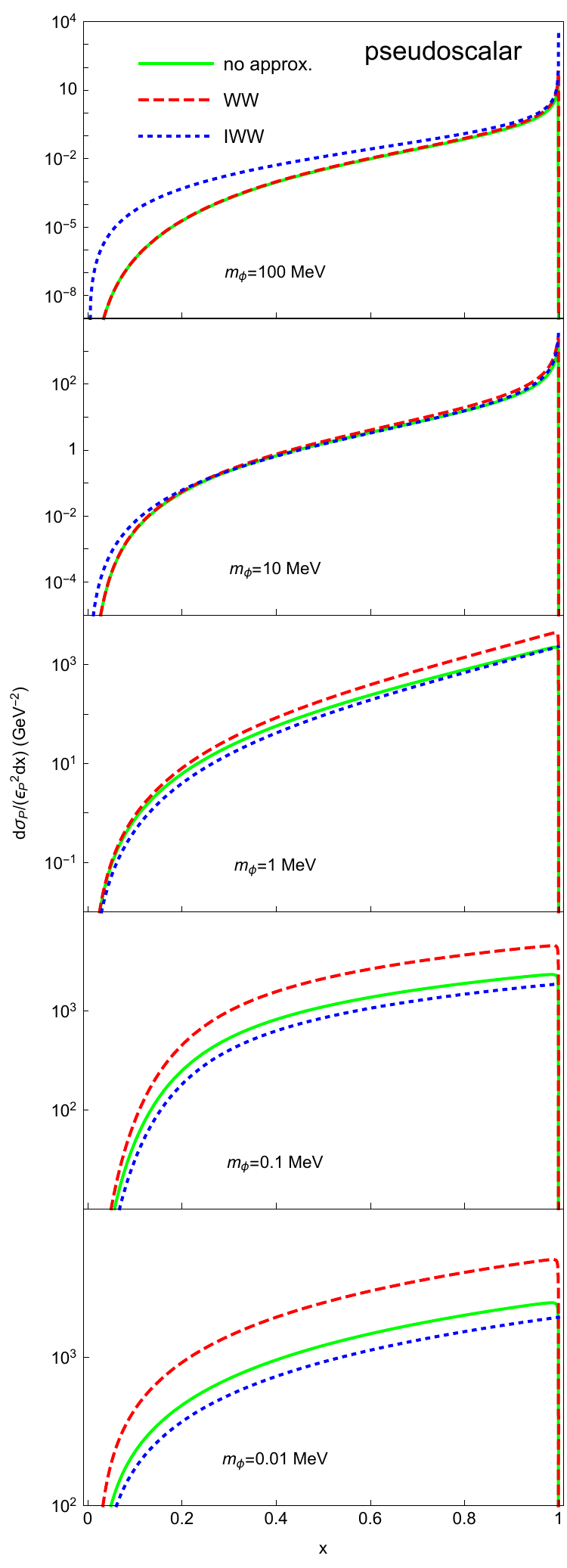}}
\subfigure[\;relative error of $d\sigma_P/(\epsilon_P^2 dx)$]{\includegraphics[scale=0.95]{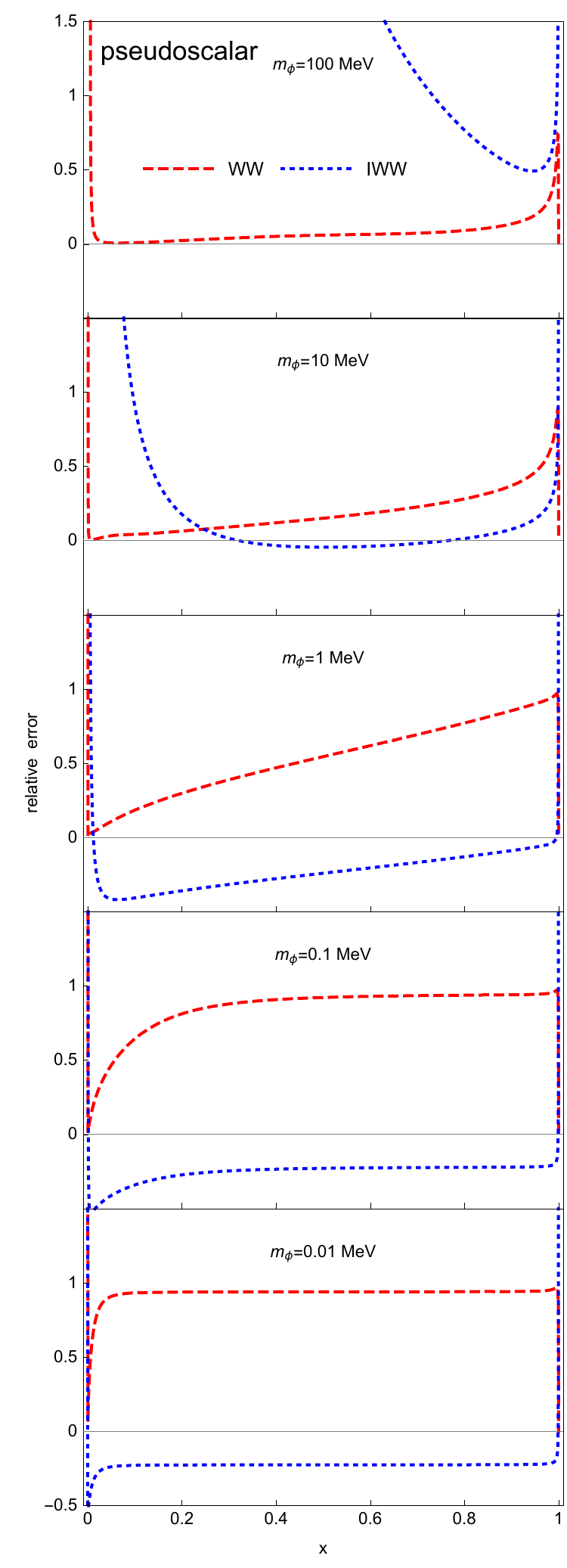}}
\caption{\label{fig:pseudoscalar_cross_section} Cross section of a pseudoscalar boson production: the solid green, dashed red, and dotted blue lines correspond to the differential cross section with no, WW, and IWW approximation. The relative error of $\mathcal{O}$ is defined by $(\mathcal{O}_{\rm approx.}-\mathcal{O}_{\rm exact})/\mathcal{O}_{\rm exact}$.}
\end{figure}

\begin{figure}
\centering
\subfigure[\;$d\sigma_V/dx$]{\includegraphics[scale=0.95]{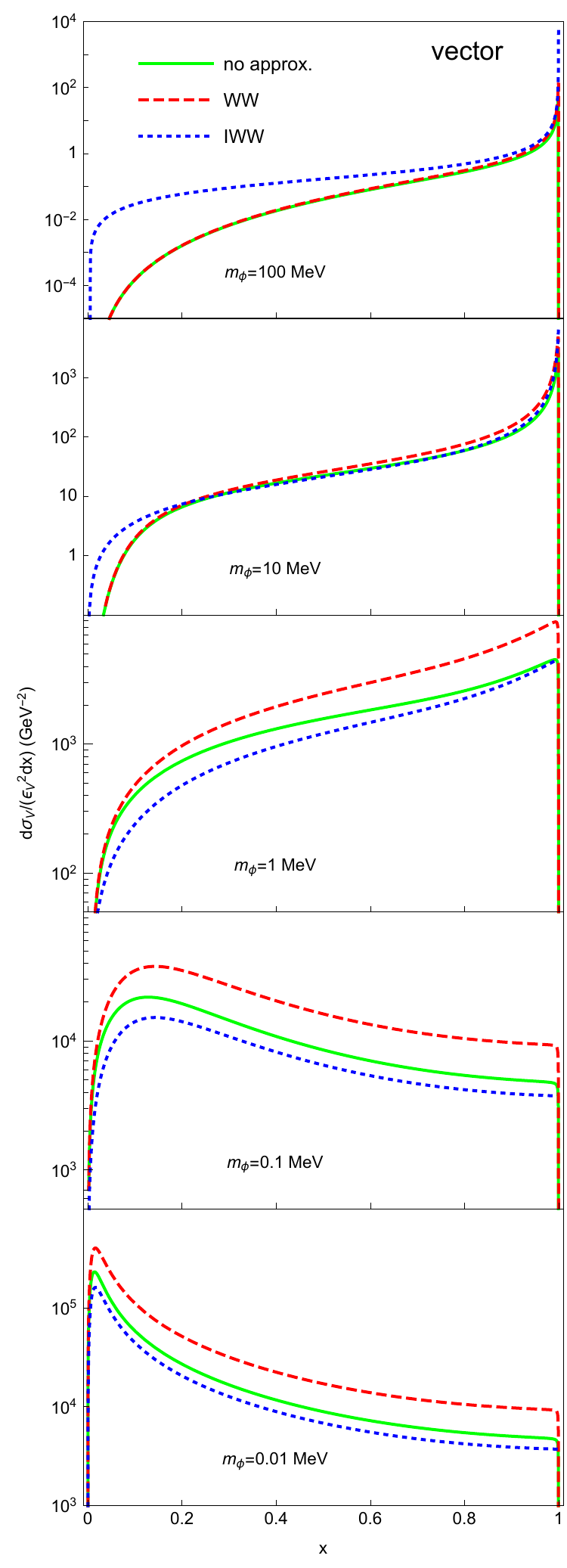}}
\subfigure[\;relative error of $d\sigma_V/(\epsilon_V^2 dx)$]{\includegraphics[scale=0.95]{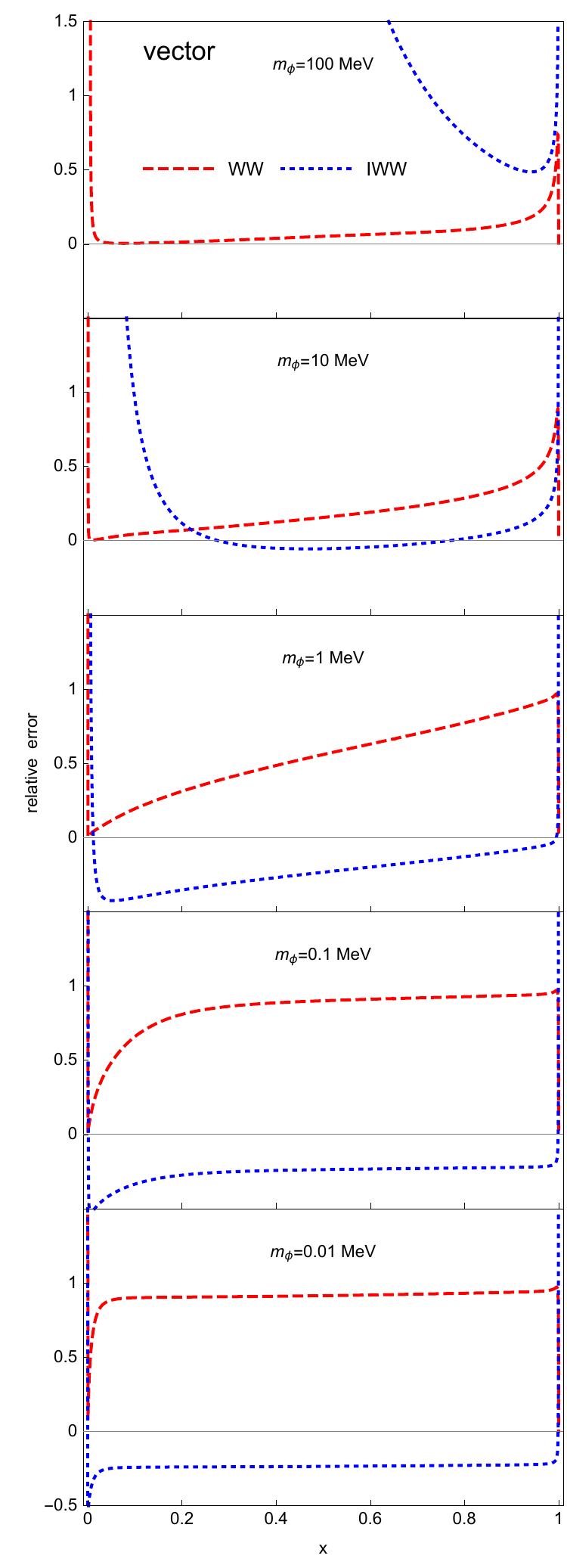}}
\caption{\label{fig:vector_cross_section} Cross section of a vector boson production: see caption of Fig. \ref{fig:pseudoscalar_cross_section} for detail.}
\end{figure}

\begin{figure}
\centering
\subfigure[\;$d\sigma_A/dx$]{\includegraphics[scale=0.95]{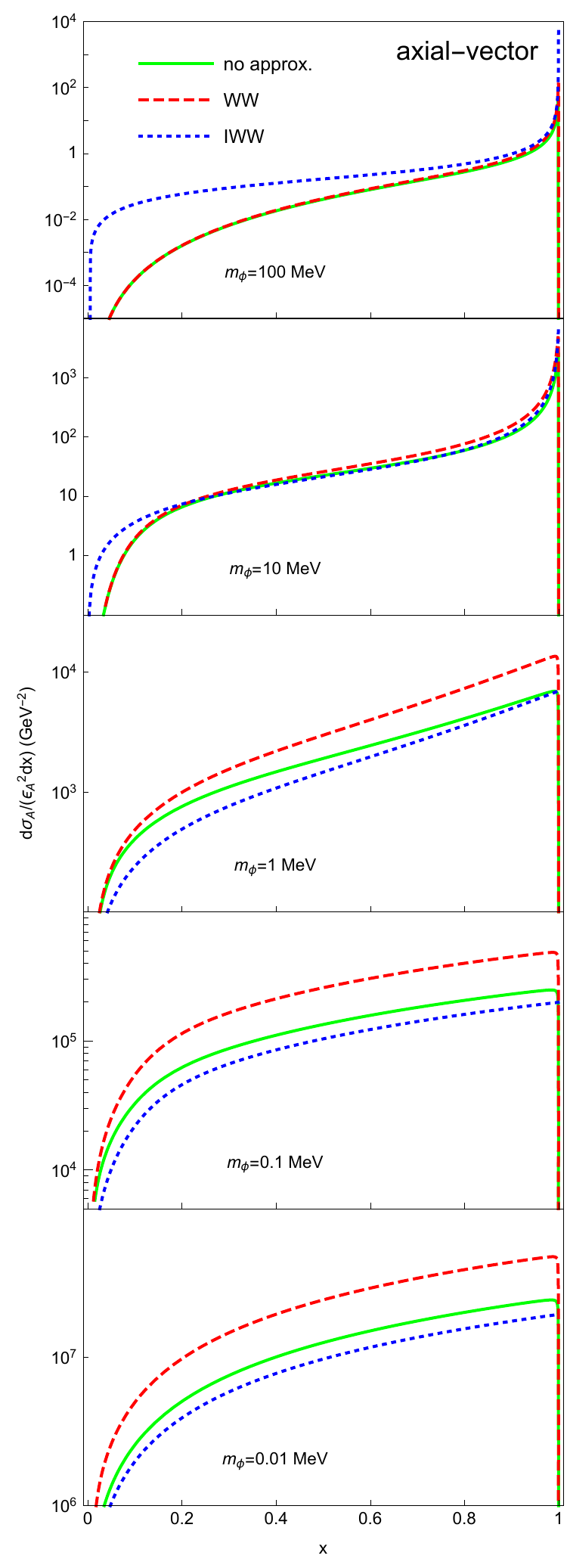}}
\subfigure[\;relative error of $d\sigma_A/(\epsilon_A^2 dx)$]{\includegraphics[scale=0.95]{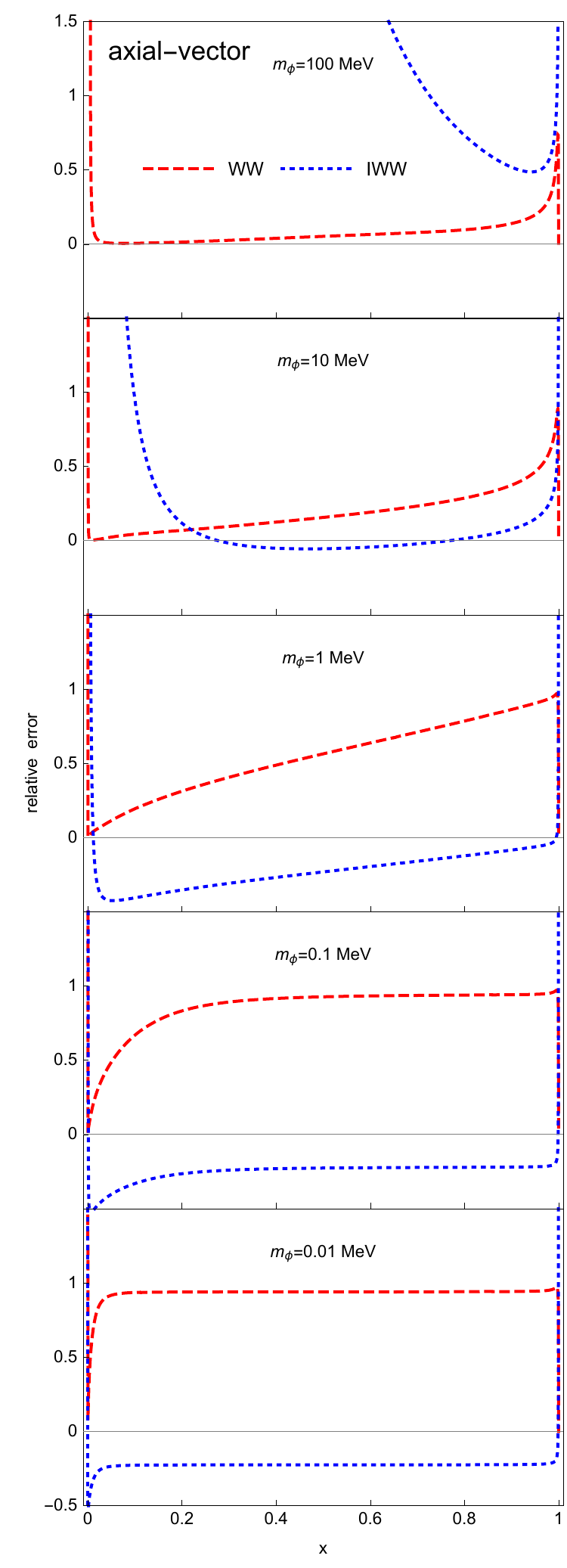}}
\caption{\label{fig:axial_vector_cross_section} Cross section of a axial-vector boson production: see caption of Fig. \ref{fig:pseudoscalar_cross_section} for detail.}
\end{figure}

To test approximations of the cross section for $\phi$ production, we examine three cases. 
\begin{enumerate}
\item The complete calculation, Eq. (\ref{eq:2 to 3}),
\begin{align}\label{eq:d sigma dx 1}
\frac{d\sigma}{dx}=\epsilon^2\frac{\alpha^3}{8M^2}\frac{|\textbf{k}|E}{|\textbf{p}|}\int_0^{\theta_{max}} d\cos\theta\frac{1}{V}\int^{t_{max}}_{t_{min}}dt\frac{F(t)^2}{t^2}\int_0^{2\pi}\frac{d\phi_q}{2\pi}\mathcal{A}^{23},
\end{align}
where $\theta_{max}$ depends on the configuration of the detector. For beam dump E137, $\theta_{max}\approx 4.4\times10^{-3}$.

\item WW: using the WW approximation, Eq. (\ref{eq:WW}),
\begin{align}\label{eq:d sigma dx 2}
\left(\frac{d\sigma}{dx}\right)_{WW}=2\epsilon^2\alpha^3|\textbf{k}|E(1-x)\int_0^{\theta_{max}} d\cos\theta\frac{\mathcal{A}^{22}_{t=t_{min}}}{\tilde{u}^2}\chi,
\end{align}
where $\theta_{max}$ is the same as the first case and $\chi=\int^{t_{max}}_{t_{min}}dt\frac{t-t_{min}}{t^2}F(t)^2$. Note that the upper and lower limits of $\chi$ depend on $x$ and $\theta$.

\item Improved WW (IWW): If the upper and lower limits of the $t$-integral in $\chi$ in Eq. (\ref{eq:d sigma dx 2}) are not sensitive to $x$ and $\theta$; i.e., the integration limit can be set to be independent of $x$ and $\theta$, we can further approximate the integration limits of $t$. Similar to the argument in Ref.~\cite{Bjorken:2009mm}, we set
\begin{align}\label{eq:tmin tmax}
t_{min}=\left(\frac{m_\phi^2}{2E}\right)^2 {\rm\; and\;\;} t_{max}=m_\phi^2+m_e^2,
\end{align}
which is valid when the production cross section is dominantly collinear with $x$ close to 1. The difference in $t_{max}$ between \cite{Bjorken:2009mm} and our approach is because we do not assume $m_\phi\gg m_e$. Therefore, we can pull $\chi$ out of the integral over $\cos\theta$. Then, changing variables from $\cos\theta$ to $\tilde{u}$ and extending the lower limit of $\tilde{u}$ to $-\infty$, 
\begin{align}\label{eq:d sigma dx 3-1}
\left(\frac{d\sigma}{dx}\right)_{IWW}&=\epsilon^2\alpha^3\chi\frac{|\textbf{k}|}{E}\frac{1-x}{x}\int^{\tilde{u}_{max}}_{-\infty}d\tilde{u}\frac{\mathcal{A}^{22}_{t=t_{min}}}{\tilde{u}^2}
\end{align}
using Eq. (\ref{eq:2 to 2 A tmin}) we have
\begin{align}\label{eq:d sigma dx 3-2}
\left(\frac{d\sigma_P}{dx}\right)_{IWW}=&\epsilon_P^2\alpha^3\chi\frac{|\textbf{k}|}{E}\frac{m_e^2x^2-2x \tilde{u}_{max}}{3\tilde{u}_{max}^2}\nonumber\\
\left(\frac{d\sigma_V}{dx}\right)_{IWW}=&2\epsilon_V^2\alpha^3\chi\frac{|\textbf{k}|}{E}\frac{m_e^2x(-2+2x+x^2)-2(3-3x+x^2)\tilde{u}_{max}}{3x\tilde{u}_{max}^2}\\
\left(\frac{d\sigma_A}{dx}\right)_{IWW}=&2\epsilon_A^2\alpha^3\chi\frac{|\textbf{k}|}{E}\left[\frac{m_e^2x(2-x)^2-2(3-3x+x^2)\tilde{u}_{max}}{3x\tilde{u}_{max}^2}+\frac{2m_e^2(1-x)}{\tilde{u}_{max}(\tilde{u}_{max}+m_e^2x)}\right]\nonumber
\end{align}
where $\tilde{u}_{max}=-m_\phi^2\frac{1-x}{x}-m_e^2 x$. We emphasize that the name ``improved" means reducing the computational time (because of one fewer integral than in the WW approximation above) and does not imply more accuracy.

\end{enumerate}

In Figs. \ref{fig:pseudoscalar_cross_section}--\ref{fig:axial_vector_cross_section}, we show the cross sections in each of the above three cases for five values of the new boson mass, setting the incoming electron beam energy to 20 GeV and the target to be aluminum. The cross sections for different bosons are different, as expected, because they have different dynamics; the relative errors with the same approximation between different bosons are similar, also as expected, because the approximation deals with phase space integral and the kinematics between different bosons are similar.

In both approximations, the cross section is of the same order of magnitude as that using the complete calculation. However, there are regions where there are ${\cal O}\left(1\right)$ relative errors. The WW approximation (dashed red lines in Figs.~\ref{fig:pseudoscalar_cross_section}--\ref{fig:axial_vector_cross_section}) can differ from the complete calculation by 100\% when $m_\phi\lesssim 1$ MeV; in the IWW case (dotted blue lines in Figs.~\ref{fig:pseudoscalar_cross_section}--\ref{fig:axial_vector_cross_section}), the approximation starts to fail when $m_\phi\gtrsim 100$ MeV.

\section{particle production}\label{sec:particle production}

There are two characteristic lengths which are crucial in beam dump experiments. The first is the decay length of the new particle in the lab frame,
\begin{align}
l_\phi=\frac{E_k}{m_\phi}\frac{1}{\Gamma_\phi},
\end{align}
where $\Gamma_\phi=\Gamma(\phi\to e^+e^-)+\Gamma(\phi\to{\rm photons})$, see Eqs. (\ref{eq:decay to electrons},\ref{eq:P decay to photons},\ref{eq:V decay to photons},\ref{eq:A decay to photons}). The new particle, after production, must decay after going through the target and shielding and before going through the detector in order to be observed. If the target is thick (much greater than a radiation length), most of the new particles will be produced in the first few radiation lengths. The production rate is approximately proportional to the probability $e^{-L_{sh}/l_\phi}(1-e^{-L_{dec}/l_\phi})$, where $L_{sh}$ is length of the target and shield and $L_{dec}$ is length for the new particle to decay into electron or photon pairs after the shield and before the detector.

The second characteristic length is the absorption length
\begin{align}
\lambda=\frac{1}{n_e\sigma_{abs}},
\end{align}
where $n_e$ is the number density of the target electrons and $\sigma_{abs}$ is the cross section of absorption process. The leading process of absorption is
\begin{align}\label{eq:absorption}
e(p)+\phi(k)\rightarrow e(p')+\gamma(q),
\end{align} 
which is related to the 2 to 2 production process Eq. (\ref{eq:2 to 2 production process}) via crossing symmetry $\tilde{s}\leftrightarrow\tilde{u}$. Since Eq. (\ref{eq:2 to 2 A}) is symmetric in $\tilde{s}\leftrightarrow\tilde{u}$, the algebraic form of amplitude squared of absorption process is the same as Eq. (\ref{eq:2 to 2 A}) but differs by a factor $c$ from summing over final state instead of averaging over initial state in Eq. (\ref{eq:2 to 2 M})
\begin{align}
\mathcal{A}^{22}_{abs}=c\mathcal{A}^{22}
\end{align}
where $c=2$ for spin-0 and $c=\frac{2}{3}$ for spin-1 particles.

The cross section of the process (\ref{eq:absorption}) is 
\begin{align}
\frac{d\sigma}{d\Omega}_{abs}&=\frac{1}{64\pi^2 m}\frac{|\mathbf{q}|}{|\mathbf{k}|}\frac{\overline{|\mathcal{M}^{22}_{abs}|^2}}{E_k+m-|\mathbf{k}|\cos\theta_\gamma}\\
\sigma_{abs}&=\epsilon^2\frac{c\pi\alpha^2}{2m|\mathbf{k}|}\int_{-1}^1 d\cos\theta_\gamma\frac{|\mathbf{q}|\mathcal{A}^{22}}{E_k+m-|\mathbf{k}|\cos\theta_\gamma}
\end{align}
where $\theta_\gamma$ is the angle between outgoing photon and incoming new particle. The new particle, after produced, must not be absorbed by the target and shield to be detected. If the target is thick (much greater than absorption length), the production rate will be approximately proportional to the probability $e^{-L_{sh}/\lambda}$.

The number of the new particles produced in terms of the cross section (without considering the absorption process) can be found in, e.g., Refs.~\cite{Bjorken:2009mm,Tsai:1986tx,Andreas:2012mt}. Using the thick target approximation and including the absorption process, we find
\begin{align}
N_\phi\approx\frac{N_eX}{M}\int_{E_{min}}^{E_0}dE\int_{x_{min}}^{x_{max}}dx\int_0^TdtI_e(E_0,E,t)\frac{d\sigma}{dx}e^{-L_{sh}\left(\frac{1}{l_\phi}+\frac{1}{\lambda}\right)}(1-e^{-L_{dec}/l_\phi})
\end{align}
where $M$ is the mass of the target atom (aluminum); $N_e$ is the number of incident electrons; $X$ is the unit radiation length of the target; $E_0$ is the incoming electron beam energy, $E_{min}=m_e+\max(m_\phi,E_{cut})$ and $x_{min}=\frac{\max(m_\phi,E_{cut})}{E}$ where $E_{cut}$ is the measured energy cutoff depending on the detectors; $x_{max}$, which is smaller  but very close to 1 ($x_{max}$ can be approximated to be $1-\frac{m_e}{E}$ if the new particle and electron initial and final state are collinear); $T=\rho L_{sh}/X$ where $\rho$ is the density of the target; $l_\phi$ is the decay length of the new particle in lab frame; $\lambda$ is the absorption length of the new particle passing through the target and shield; $I_e$, derived in Ref.~\cite{Tsai:1966js}, is the energy distribution of the electrons after passing through a medium of $t$ radiation length  
\begin{align}
I_e(E_0,E,t)=\frac{\left(\ln\frac{E_0}{E}\right)^{bt-1}}{E_0\Gamma(bt)}
\end{align}
where $\Gamma$ is the gamma function and $b=4/3$. For beam dump E137 which we take as our prototypical setup, $E_0=20$ GeV and $E_{cut}=2$ GeV; $N_e=1.87\times 10^{{20}}$; $L_{sh}=179$ m and $L_{dec}=204$ m. The experiment has a null result which translates to 95\% C.L. of $N_\phi$ to be 3 events.

\section{exclusion plots}\label{sec:exclusion plots}

\begin{figure}
\centering
\includegraphics[scale=2]{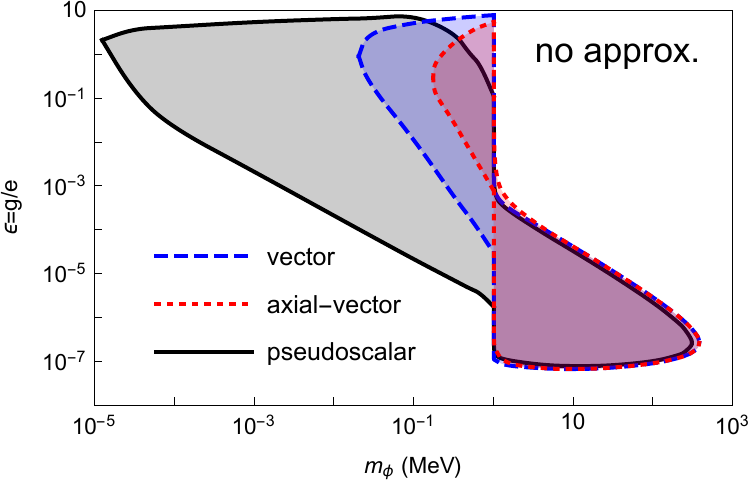}
\caption{\label{fig:E137all} Exclusion (shaded region) plot for pseudoscalar (black), vector (dashed blue), and axial-vector (dotted red) bosons without approximation using the beam dump experiment E137.}
\end{figure}

\begin{figure}
\centering
\subfigure[\;exclusion plot (mass in linear scale)]{\includegraphics[scale=1]{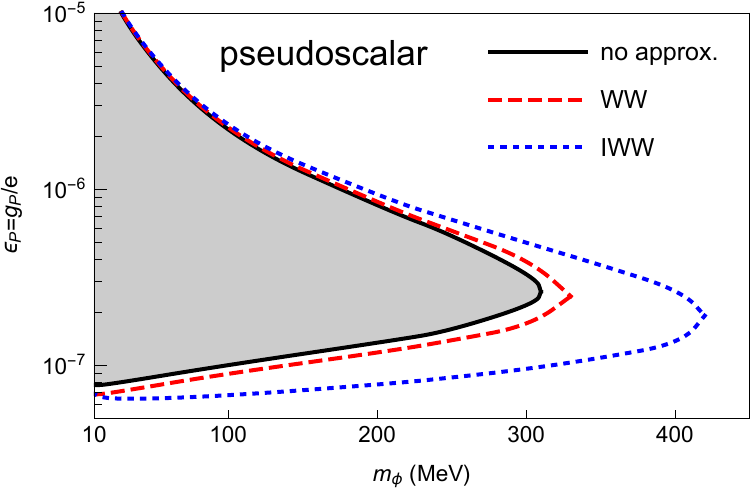}}
\subfigure[\;relative error of exclusion boundary]{\includegraphics[scale=1]{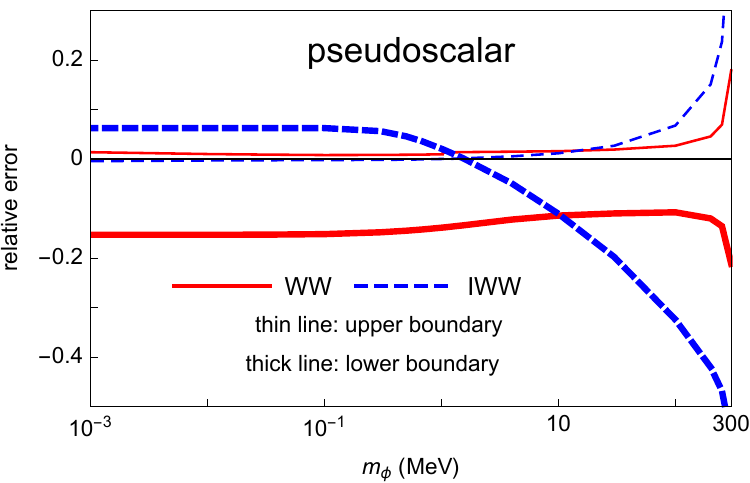}}
\caption{\label{fig:E137P} Exclusion (shaded region) plot for $\epsilon_P$ using the beam dump experiment E137: (a) The solid black, dashed red, and dotted blue lines correspond to using the differential cross section with no, WW, and IWW approximation. (b) The solid red and dashed blue lines correspond to the relative error of the exclusion boundary of (a) for a fixed value of $m_\phi$ with WW and IWW approximation. The relative error of $\mathcal{O}$ is defined by $(\mathcal{O}_{\rm approx.}-\mathcal{O}_{\rm exact})/\mathcal{O}_{\rm exact}$. The thin and thick lines correspond to the upper and lower boundaries of the exclusion plot.}
\end{figure}

\begin{figure}
\centering
\subfigure[\;exclusion plot (mass in linear scale)]{\includegraphics[scale=1]{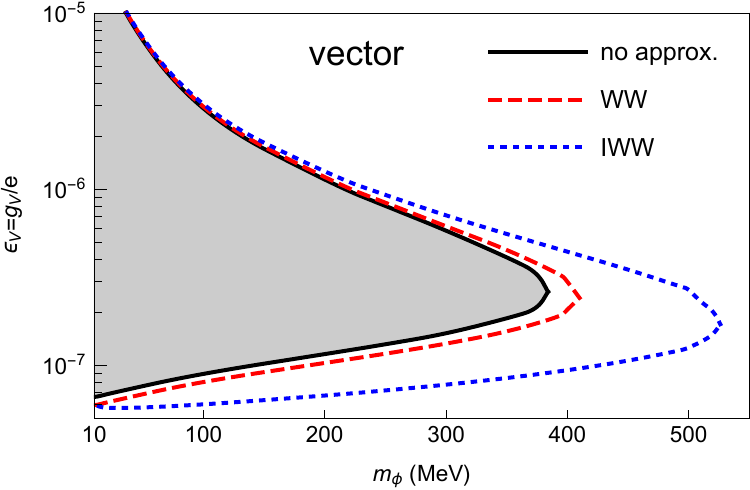}}
\subfigure[\;relative error of exclusion boundary]{\includegraphics[scale=1]{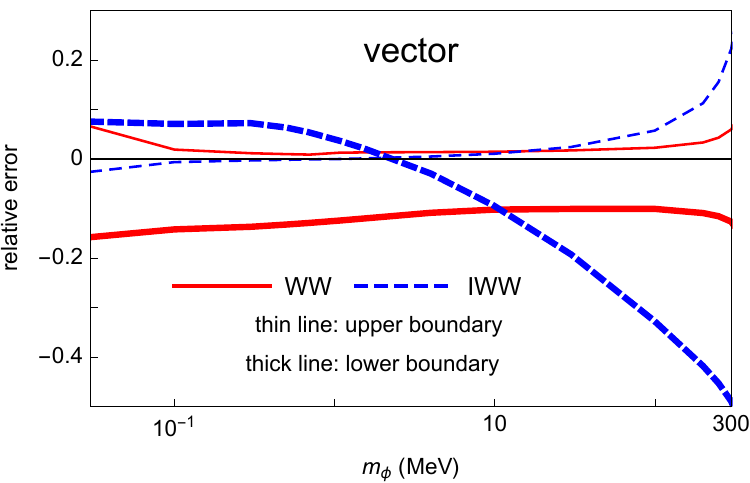}}
\caption{\label{fig:E137V} Exclusion (shaded region) plot for $\epsilon_V$: see caption of Fig. \ref{fig:E137P} for detail.}
\end{figure}

\begin{figure}
\centering
\subfigure[\;exclusion plot (mass in linear scale)]{\includegraphics[scale=1]{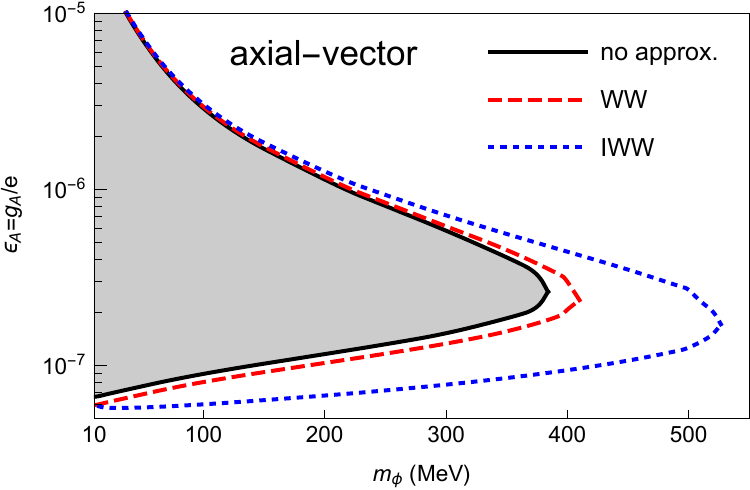}}
\subfigure[\;relative error of exclusion boundary]{\includegraphics[scale=1]{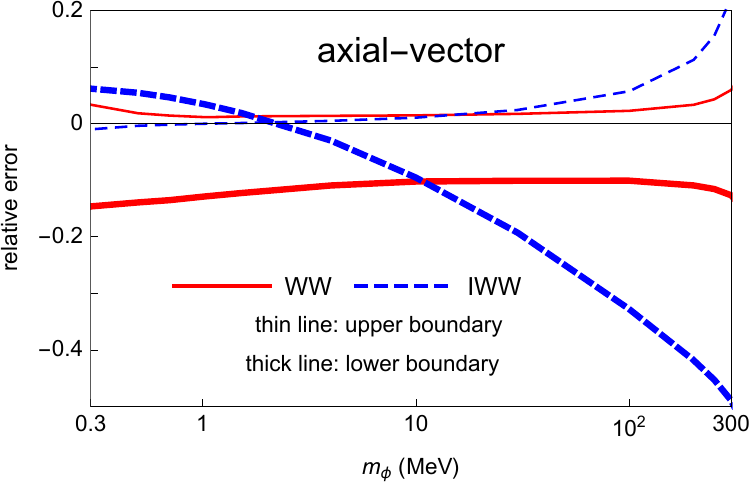}}
\caption{\label{fig:E137A} Exclusion (shaded region) plot for $\epsilon_A$: see caption of Fig. \ref{fig:E137P} for detail.}
\end{figure}

In Fig.~\ref{fig:E137all}, using Eq. (\ref{eq:d sigma dx 1}), we show regions of coupling and mass excluded by the lack of a signal at E137 for different bosons. In the region where $m_\phi>2m_e$ the exclusion plots are similar with each other, however, in the region where $m_\phi<2m_e$ the exclusion plots are very different because the the decay widths for different bosons are fundamentally different. 

In Figs.~\ref{fig:E137P}--\ref{fig:E137A}, using Eq. (\ref{eq:d sigma dx 1},\ref{eq:d sigma dx 2},\ref{eq:d sigma dx 3-2}), we show the exclusion regions using the three different ways to calculate the differential cross section. Because of the exponential factor from decay and absorption lengths, the error in the exclusion plot due to making approximations to the cross section is smaller along the upper boundary, which is mainly determined by whether $\phi$ lives long enough to make it to the detector. With the WW approximation, the 100\% error in cross section causes an error of less than 20\% along the lower boundary, and in a log-log plot across several scales, a 20\% error is almost indistinguishable by eyesight. On the other hand, with the IWW approximation, the difference is clearly visible when $m_\phi\gtrsim 100$ MeV. In the region where $m_\phi>2m_e$, the relative errors of the exclusion plots boundary for different bosons are similar based on the same reason which causes the similar relative errors of the cross section: the approximations deal with the phase space integral and the kinematics for different bosons are similar.

In Fig.~\ref{fig:E137all}, we see that the absorption process, Eq. (\ref{eq:absorption}), cuts off the exclusion plot around $\epsilon\sim\mathcal{O}(1)$ where the coupling of $\phi$ to electrons is of same order of the electromagnetic coupling. Therefore, in this region, there is another significant process to consider for beam dump experiments. This is the trapping process due to the rescattering
\begin{align}
e(p)+\phi(k)\rightarrow e(p')+\phi(k').
\end{align} 
The trapping process is expected to be as important as the absorption process in this example (new bosons and beam dump E137), and also cuts off the exclusion plot around $\epsilon\sim\mathcal{O}(1)$. However, in Fig.~\ref{fig:E137all} the region where $\epsilon\sim\mathcal{O}(1)$ has been excluded by other experiments, such as electron $g-2$ \cite{Pospelov:2008zw,Bouchendira:2010es} and hydrogen Lamb shift \cite{Eides:2000xc}, which are discussed in Ref.~\cite{Liu:2016qwd} as well as astrophysical processes \cite{Essig:2013lka}. Therefore we do not include the trapping process, but it might be crucial for other experiments.

\section{discussion}\label{sec:discussion}
In the region where $m_\phi>2m_e$, while the production amplitude, decay length, and the absorption length can differ in detail for particles with different quantum numbers, they are qualitatively similar. The approximations that we have examined deal with the phase space integral and coupling to electromagnetism of the target nucleus. Therefore, as we expected, the exclusion plots for different bosons are similar. On the other hand, where $m_\phi<2m_e$, the decay channels, which are very different for different bosons, result in very different exclusion regions. New results for vector decaying to 3 photons, Eq. (\ref{eq:V decay to photons}), and axial-vector decaying to 4 photons, Eq. (\ref{eq:A decay to photons}), are presented.

Including a coupling to the muon may change the situation for $m_\phi>2m_\mu$~\cite{Liu:2016qwd} due to the opening of a new channel with typically a substantial partial width. A study of the production of vector particles in electron beam dumps that deals with some of the issues we have addressed can be found in Ref.~\cite{Beranek:2013yqa}.

There are some other beam dump experiments using a Cherenkov detector, such as E141 \cite{Riordan:1987aw} and Orsay \cite{Davier:1989wz}. Therefore, their exclusion plots do not extend to the region where $m_\phi<2m_e$. We show the results of the beam dump experiments E141 and Orsay for the scalar boson in Ref.~\cite{Liu:2016qwd}.

We need to consider the LPM effect \cite{Landau:1953um,Landau:1953gr,Migdal:1956tc,Anthony:1995fs} which suppresses particle production cross section below a certain (produced particle) energy. For E137, this energy is about 12 MeV which is much smaller than the energy cutoff of the detector. Therefore we do not consider the LPM effect in our discussion. However, for other experiments (depending on the apparatus), the LPM effect may need to be taken into account.

In this work, we present a complete analysis of beam dump experiments. We show that a brute-force analytical calculation is possible. Software exists using Monte-Carlo simulations, such as \textsc{MadGraph/MadEvent}~\cite{Alwall:2007st} as used in, e.g.,~\cite{Essig:2010xa}, that can calculate the cross section without using approximations. Our work can be used as a consistency check for Monte-Carlo simulations. We show that using the WW approximation can be trusted to an order of magnitude in cross sections and exclusion plots. Additionally our work  allows us to understand the errors introduced by the various common approximations. In certain regions of parameter space different errors partially cancel against each other, leading to results that are accidentally sometimes more accurate than one might be expected. However, in our previous work, we illustrated with several pseudoexperiments that in the event of a nonzero signal, a complete calculation is necessary. This work could be useful given the possibility of future beam dump experiments or beam dump like experiments~\cite{future}.

\section*{acknowledgement}
We acknowledge M. McKeen and A. E. Nelson for invaluable discussions and suggestions. The work of G. A. M. and Y.-S. L. was supported by the U. S. Department of Energy Office of Science, Office of Nuclear Physics under Award Number DE-FG02-97ER-41014.

\end{document}